\documentclass[twocolumn,aps, prl,showpacs]{revtex4}

\usepackage{graphics}% Include figure files
\usepackage{graphicx}% Include figure files
\usepackage{dcolumn}% Align table columns on decimal point
\usepackage{bm}% bold math
\usepackage{amsmath,amssymb}
\usepackage{color}
\usepackage{soul,xcolor}
\setstcolor{red}

\begin{document}

\title{How fast is a twisted photon?}
\author{Thomas Roger$^{1}$, Ashley Lyons$^{1}$, Niclas Westerberg$^{1}$, Stefano Vezzoli$^{1}$, Calum Maitland$^{1}$, Jonathan Leach$^{1}$, Miles Padgett$^{2}$, Daniele Faccio$^{1}$}
\email{t.w.roger@hw.ac.uk, d.faccio@hw.ac.uk}
\affiliation{$^{1}$School of Engineering and Physical Sciences, Heriot-Watt University, Edinburgh, EH14 4AS, UK\\
$^{2}$Physics $\&$ Astronomy, Kelvin Building, Glasgow, G12 8QQ, UK}

\date{\today}
\begin{abstract}
Recent measurements have highlighted that spatially shaped photons travel slower than $c$, the speed of monochromatic, plane waves in vacuum. Here we investigate the \emph{intrinsic} delay introduced by `twisting' a photon, i.e. by introducing orbital angular momentum (OAM). In order to do this we use a Hong-Ou-Mandel interferometer to measure the change in delay of single photons when we introduce OAM on a ring-shaped beam that is imaged through a focusing telescope. Our findings show that when all other parameters are held constant the addition of OAM reduces the delay (accelerates) with respect to the same beam with no OAM. We support our results using a theoretical method to calculate the group velocity and gain an intuitive understanding of the measured OAM acceleration by considering a geometrical ray-tracing approach.
\end{abstract}
\maketitle
\begin{figure}[t!]
\centering{
\includegraphics[width = 0.4\textwidth]{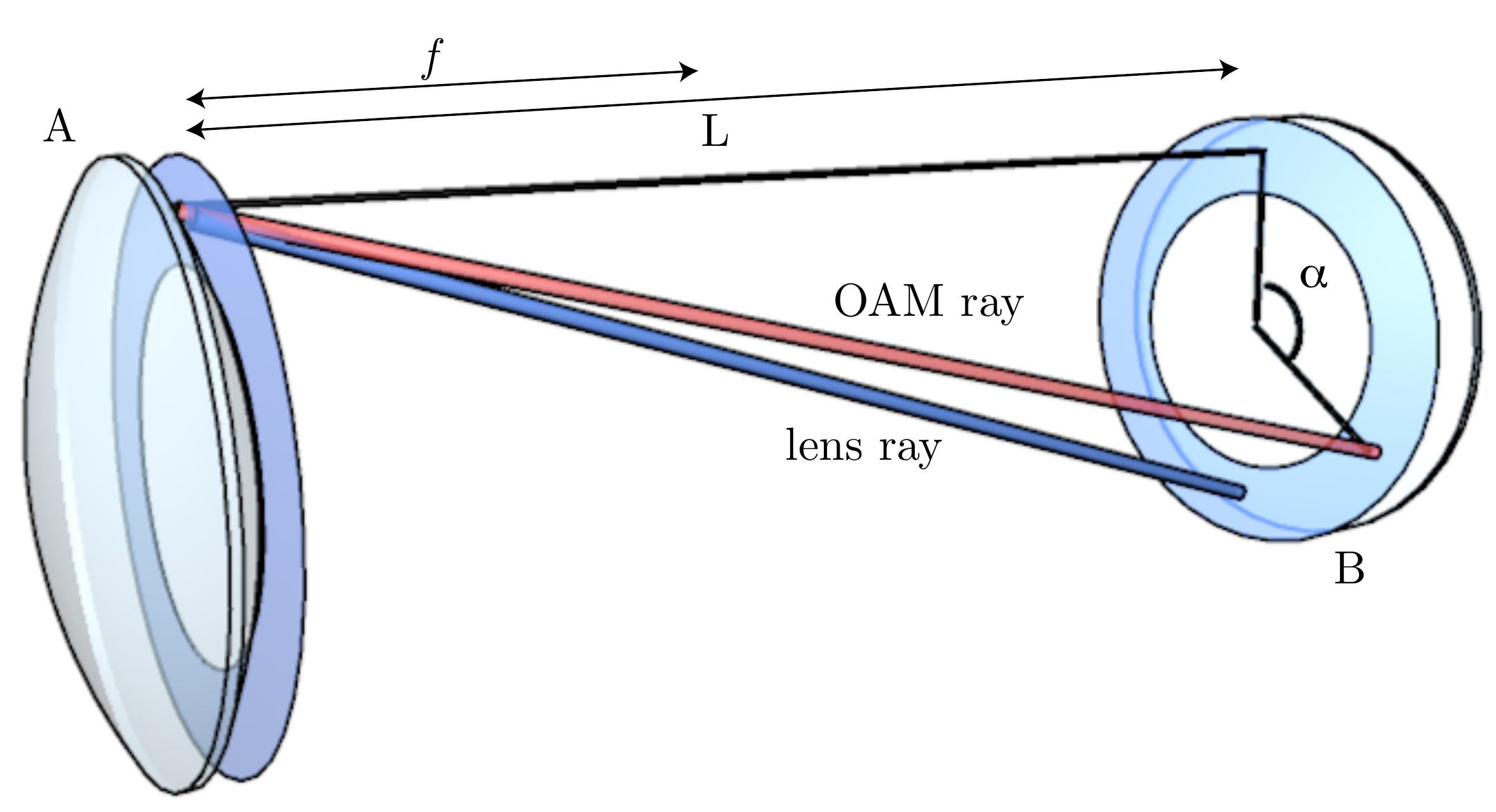}
\caption{Ray picture of the plane wave trajectories in the case of a fixed-diameter ring shaped beam passing through the focus of a telescope formed between lenses placed at positions A and B (spaced by distance $L = 2f$). In the zero-OAM $m=0$ case (indicated as `lens ray') images are inverted, corresponding to a rotation of the ray by an angle $\alpha = 180^{\circ}$. Introducing OAM  (indicated as `OAM ray') necessarily leads to a smaller rotation angle  $\alpha$ and thus to a shorter path length covered by the respective planes waves, i.e. to an acceleration of the photon.}
\label{fig1}}
\end{figure}

\noindent {\bf Introduction.} The propagation velocity of light in vacuum is known according to relativity to be constant, $c$. However, this is strictly true only for monochromatic plane waves: any deviation from the plane wave constraints can lead to a propagation velocity different from $c$. Even the simplest of light beams, such as a focused Gaussian beam, exhibits regions of superluminal propagation \cite{horvath}. A whole range of more complicated beams that are appropriately shaped in space, time or space-and-time have been identified that exhibit intensity peaks and structures that are subluminal, superluminal or accelerating \cite{mugnai,milchberg,saari1,saari2,faccio1,airy}. \\
A slightly more fundamental issue has recently been addressed, namely the propagation time of a single photon from one point (the `source') to another (the `detector'), highlighting that beam structure will modify the propagation time. For instance, one can measure a longer transit time for a photon with a constant non-zero transverse wavevector, such as a Bessel mode, when compared to a non-structured (plane-wave) photon \cite{giovannini}. The speed of a photon structured in the transverse dimension propagating in vacuum,  can be calculated as: 
\begin{equation}
\langle v^{(z)}_{g} \rangle = \left\langle \left(\frac{\partial^2 \phi(\mathbf{r})}{\partial_\omega\partial_z}\right)^{-1}_{\omega_0} \right\rangle,
\label{eq:vg}
 \end{equation}
where $\partial_\omega$ and $\partial_z$ indicate the gradient with respect to the photon carrier frequency $\omega$ and propagation direction $z$ respectively and $\phi(\mathbf{r})$ represents the phase front in the cylindrical coordinates $\mathbf{r}=(r,\phi,z)$ \cite{born}. Here the $\langle{...}\rangle$ denotes the spatial average over the transverse structure, using the normalised field amplitude as a weight. In the case of a plane wave, $\partial_z\phi(\mathbf{r})=k_0$, where $k_0=2\pi/\lambda$ is the vacuum wave-vector of the photon with wavelength $\lambda$. This plane wave provides the standard result that in vacuum the group velocity is equal to the speed of light $c=\omega/k_0$. However, as soon as there is a spatial structure of any kind on the transverse photon profile,  $\partial_z\phi(\mathbf{r})$ is a non-trivial function of $\omega$ such that the speed along the propagation direction will deviate from $c$. An intuitive understanding of this deviation is based on the simple observation that $c$ is the speed of a monochromatic plane wave (as can be seen upon inspection of the Maxwell equations) and that in a structured photon, the constituent plane waves are propagating at non-zero angles with respect to the propagation axis.  The travel time of a photon from one point `A' to another point `B' is longer due to  plane wave components travelling at an angle with respect to the axis connecting A to B (and hence travel a longer distance, see Fig.~\ref{fig1}). It is therefore intuitively clear that the propagation time of a focused (e.g. imaged from the the source to the detector) ring-shaped photon will increase as the ring-radius is increased. This delay of a ring-shaped photon is also well described within the framework of a simple ray-tracing theory \cite{giovannini}. \\
The propagation of light with an azimuthal phase gradient i.e. orbital angular momentum (OAM), was also recently examined \cite{bouchard, bareza}. In this case, one must allow for the fact that these beams also have a ring-shaped spatial profile and that for increasing OAM, the ring radius also naturally increases for the typical beam types encountered in experiments e.g. Laguerre-Gauss or Hypergeometric beams. The introduction of OAM will therefore lead to the observation of a delay for increasing OAM due the increasing ring size \cite{bouchard}. The effective change in velocity of beams carrying OAM may have consequences in a number of quantum technologies, such as quantum key distribution \cite{mafu, mirhosseini}, which use OAM due to it's inherent high dimensionality \cite{leach, agnew, dada}. For example, accurate timing of signals used in quantum communications may be affected if the velocity of the photons is calculated incorrectly \cite{gibson}. Furthermore, measurements of photons in cosmology could be subject to discrepancies in arrival time if proper account is not taken for these subtle effects \cite{chelkowski}.\\
Here we address the fundamental question of the intrinsic effect of OAM on the propagation time of a single photon. This study requires separating the effect of increasing OAM from the effect of the associated increase in ring radius in naturally occurring OAM modes. Our measurements show that, rather counterintuitively, OAM \emph{increases} the velocity of the photon when compared to a photon that has exactly the same spatial profile but is carrying zero OAM. These findings are supported by theoretical models and ray-tracing considerations.\\

\noindent {\bf Theory.} We wish to study the propagation velocity of photons with a fixed spatial profile, specifically those with a ring shaped intensity distribution. A useful set of modes to consider is the Bessel-Gauss set \cite{bgmodes}. These beams exhibit an extended depth of field and an annular intensity distribution in the far field \cite{bagini, vasilyeu,bandres}. Following Ref.~\cite{giovannini}, we can calculate the $z$-average velocity of a photon:
\begin{equation}
\overline{v_g} = L \left(\int_{z_1}^{z_2} dz \; \langle v^{(z)}_{g} \rangle \right)^{-1},
\end{equation}
where $L = z_2-z_1$ and where $z_1$ and $z_2$ indicated the start (source) and end (detector) positions of the photon path. The time taken for a photon to travel $L$ is thus $t = L/\overline{v_g}$, and the distance between a structured photon and a plane wave that travels at constant speed $c$ over a fixed time is given by
%\begin{widetext}
\begin{eqnarray}
\delta z &= & ct - L = \nonumber\\
&=& \left[\frac{\partial}{\partial k} \bigg(\arg\langle \psi (\mathbf{r}, k_0)| \psi(\mathbf{r},k)\rangle\bigg)\bigg|_{k_0}-z\right]_{z_1}^{z_2}.
\label{delay1}
\end{eqnarray}
where $\arg$ is the complex argument, and $|\psi(\mathbf{r},k)\rangle$ is the normalised field amplitude such that the $\langle \psi | \psi \rangle$ denotes the overlap integral in the transverse plane $(r,\phi)$.
%\end{widetext}
In the case of the paraxial wave approximation, this simplifies to \cite{giovannini}:
\begin{equation}
\overline{v_g} = \frac{c}{1+\frac{\langle {\bf k}_{\perp}^2\rangle_{z_1}}{2k_0^2}} \approx c \left(1-\frac{\langle\mathbf{k}_{\perp}^2\rangle_{z_1}}{2k_0^2}\right)
\end{equation}
and
\begin{equation}
\delta z = \frac{L}{2k_0^2}\langle {\bf k}_{\perp}^2\rangle_{z_1},
\label{eq:delay}
\end{equation}
where $\langle{...}\rangle_{z_1}$ denotes average over the normalised field amplitude at the source $z = z_1$ and ${\bf k}_{\perp}$ is the transverse wavevector.
This analysis provides a rigorous estimation of the change in propagation velocity of the structured photons we use in experiments. However, a more intuitive understanding of the influence of OAM on the speed of a photon may be obtained within the context of a geometrical ray description. In this description we consider a Laguerre-Gauss mode and therefore this approximation is valid for the mode overlap of the Bessel-Gauss with a pure LG mode with no radial momentum, which we show later gives a good approximation to the result predicted by the complete theory. The delay is measured by imaging the photon from the `source' to the `detector', which are placed in the object and image planes of a non-magnifying telescope. A Laguerre-Gauss LG$_{0,0}$ mode (a Gaussian mode with no OAM) will be spatially inverted as it travels through the beam waist, i.e. images will appear inverted at the telescope imaging plane. In Fig.~\ref{fig1}(a) we show the Poynting vector for this mode (indicated as ``lens ray'') that goes though an effective rotation in the transverse plane of $\alpha = 180^{\circ}$. A Laguerre-Gauss LG$_{m,0}$ mode with OAM integer $m$, behaves somewhat differently: the Poynting vector no longer rotates by  $180^{\circ}$. Rather, the vector rotates by an angle that is smaller in absolute value and either in a clockwise or counter-clockwise direction depending on the sign of OAM \cite{arlt}. \\
Using this geometrical insight it is therefore possible to estimate the photon path difference between a ring shaped beam with and without OAM (but with fixed ring radius):
\begin{equation}
\delta z_{\text{Ray Tracing}} = \left( \cfrac{1}{2}- \cfrac{1-\cos\alpha}{4} \right)\cfrac{D^2}{L}
\label{eq:dDelta}
\end{equation}
where $L$ is the distance between the two lenses, $D$ is the diameter of the intensity ring. As can be seen from Eq.~\eqref{eq:dDelta} based on ray-tracing, any OAM will actually \emph{advance} the photon with respect to a beam with same intensity profile and $m =0$. In other words, the longest possible trajectory that a light ray can take is given by image inversion, $\alpha=180^{\circ}$, corresponding to what is expected for an $m =0$ beam. If one adds OAM, the rays will be skewed yet still remain rays, i.e. straight lines that can therefore only connect the object to the image plane via trajectories that are shortened. These shortened trajectories imply a faster transit time for photons that carry OAM with respect to the photons that carry none.\\
In experiments, the actual value of the angle $\alpha$ may depend on the specific details of the lens system and beam propagation: this should therefore be determined phenomenologically, as shown below.\\
\begin{figure}[t!]
\centering{
\includegraphics[width = 0.5\textwidth]{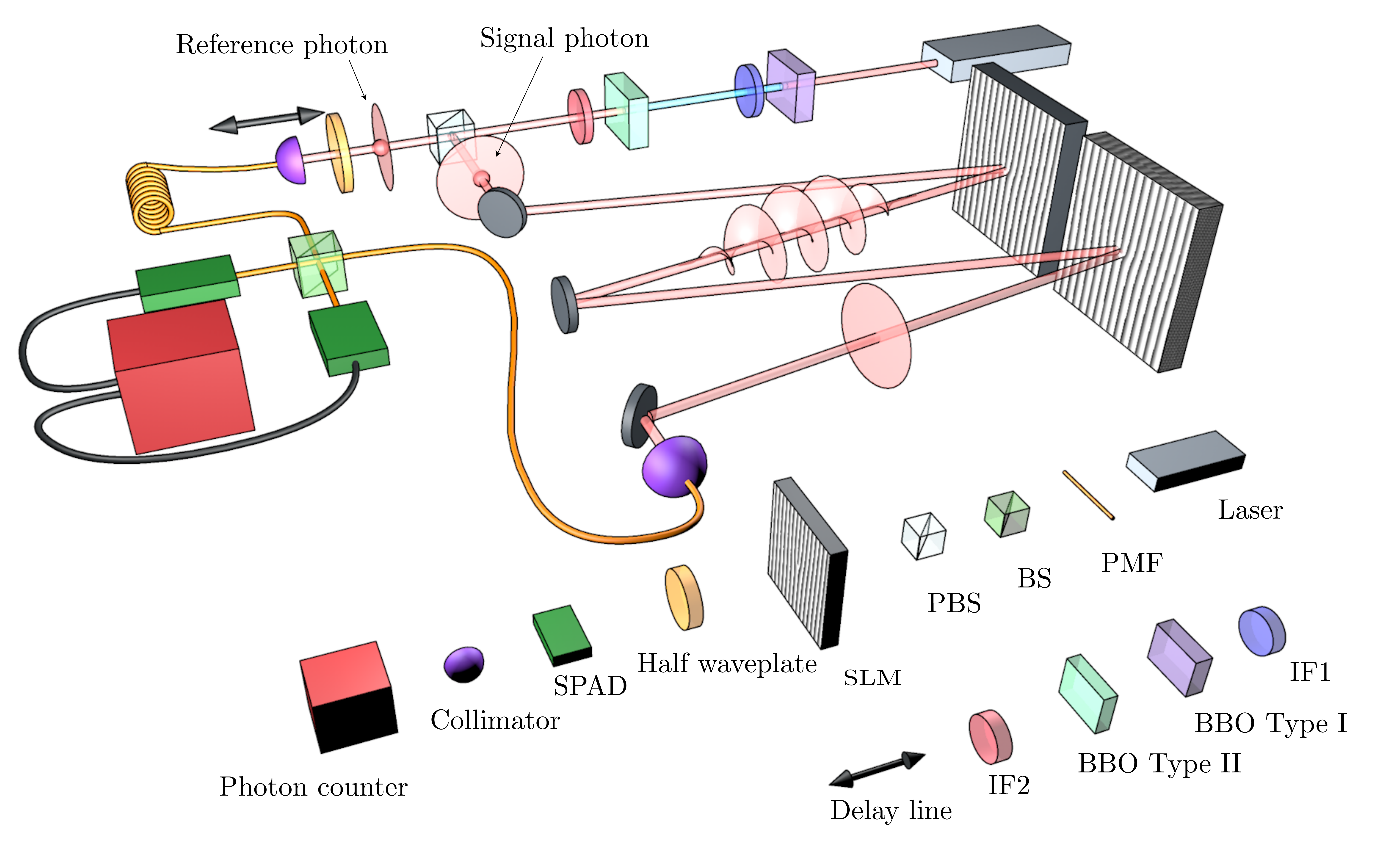}
\caption{Experimental setup to study the path length travelled by a structured photon. Pulses of light with central wavelength 404 nm are produced by frequency doubling a ti:sapph laser in a type I beta-barium borate (BBO) crystal.  Photon pairs at 808 nm are generated in a type II BBO crystal. The photons are orthogonally polarised and are split on a polarising beam splitter. One photon is spatially structured via an SLM, while the second photon (reference) is coupled to polarisation maintaining fiber (PMF) and traverses a path length that is controlled by with a stepper motor (delay line). The structured photon phase structure is returned to a plane wave by a second SLM and coupled into a PMF and recombined with the reference photon on a 50:50 beam splitter where the photons undergo HOM interference. Single photon detectors are used to coincidence count the output photons. The position of the HOM interference dip measures the transit time of the signal photon. }
\label{setup}}
\end{figure}
\noindent {\bf Experiments.} Indistinguishable photon pairs are generated via spontaneous parametric down conversion from a nonlinear crystal pumped by 100 fs, 404 nm wavelength pulses, provided by a frequency doubled 80 MHz, Ti:Sapph oscillator. The experimental arrangement is thereafter the same as that described in Ref.~\cite{giovannini}, see Fig.~\ref{setup} for details. The two photons are separated, one photon is used a plane-wave reference, the other is sent through a pair of SLMs. The first SLM imprints the desired phase structure while the second applies the inverse phase to return the photon to a plane wave mode. In all experiments (aside from plane wave propagation) a focusing lens phase is imprinted on each of the SLMs to form a 4$f$ telescope between the input and output. The signal and reference photons are recombined at a 50:50 fibre beamsplitter at which point they undergo Hong-Ou-Mandel interference \cite{hom} and produce the characteristic HOM dip as the variable path of the reference photon is tuned. The photons are detected via two single photon avalanche detectors (SPADs, Excelitas SPCM-14) and counted in coincidence via an event timing module (PicoQuant HydraHarp 400). The position of the HOM dip minimum is used to determine the transit time of the structured  photons. For each delay position of the reference photon  (controlled with a stepper motor) along the HOM dip, we iteratively imprint the various ring shaped holograms onto the SLMs and integrate the coincidence counts for 2 seconds. The measurements are then repeated 90 times and we take the final average value of the HOM dip minimum and standard deviation error. \\
We performed experiments with a ring-shaped mask applied to the beam, where we choose the size of the mask to match the natural size of a LG beam after propagation to the far field, given by $D(m)=2w\sqrt{|m|/2}$. The phase ($2\pi m$ in the range $1<m<8$) and amplitude mask were imposed onto the beam using the first SLM along with a focusing lens phase with focal length $f=300$ mm.  We also measure, for reference, the HOM dip position for a plane wave i.e. no structuring of the transverse phase. Figure~\ref{fig:raw} shows the data for each aperture radius for the case with OAM (red squares) and without OAM (blue circles). 

The solid line shows the prediction based on the theoretical analysis outlined above, where we have calculated Eq.~\eqref{eq:delay} for the specific Bessel-Gauss beams used in the experiments. We show this analysis for the case with OAM (solid red line) and without OAM (dashed blue line). We can see from these results that both experiments and theory show that photons with OAM suffer less delay than the same (spatial-profile) photon with no OAM.\\\\
\begin{figure}[t!]
\centering{
\includegraphics[width = 0.5\textwidth]{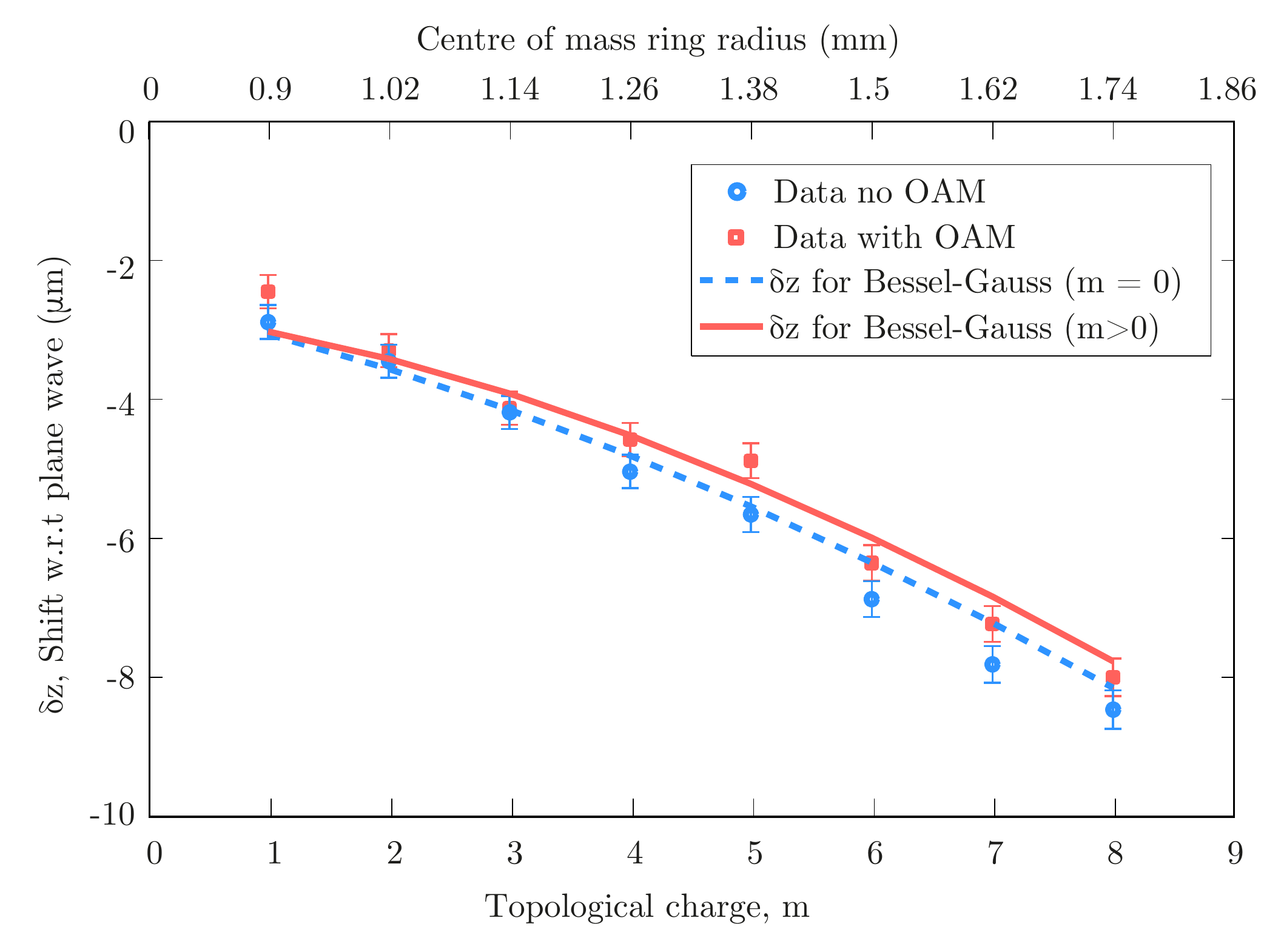}
\caption{Experimental results. Photon path delay, measured relative to a Gaussian ($m=0$) beam, for photons with a ring aperture for the case with (red squares) and without (blue circles) OAM. The ring diameter of the photons is chosen according to the far-field ring size of a HyGG beam carrying an topological charge $m$, which to good precision follows the prediction based on LG modes, $D(m)=2w\sqrt{|m|/2}$. Each ring size can be compared to the plane-wave ring (i.e. without OAM). The solid line shows the theoretical prediction based on Eq.~\eqref{eq:delay}, as described in the text.}
\label{fig:raw}}
\end{figure}
\begin{figure}[t!]
\centering{
\includegraphics[width = 0.5\textwidth]{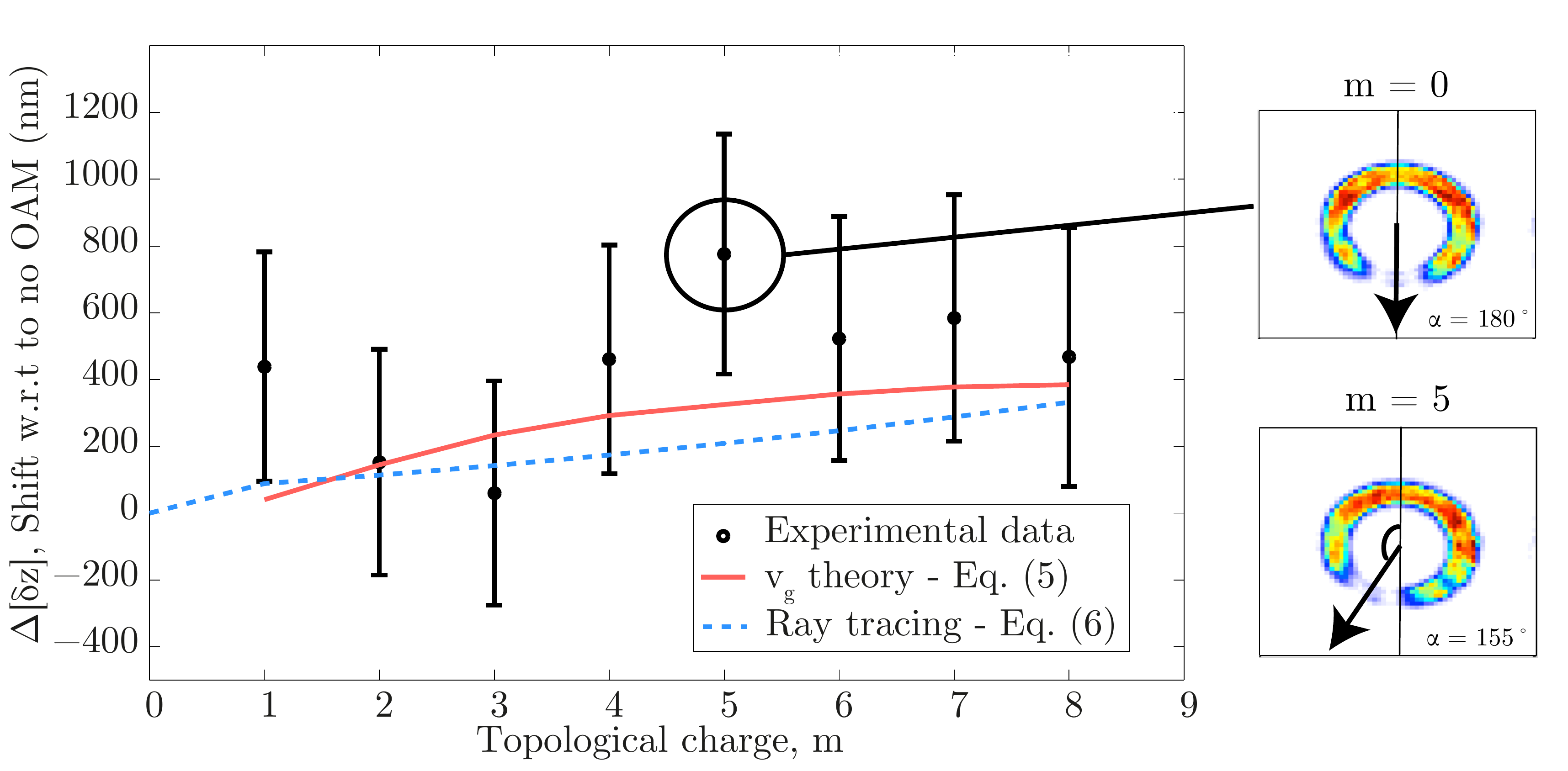}\\
\caption{Effect of adding OAM to beams with a ring aperture. We show the difference in path length between ring shaped plane wave structured photons and the same ring shaped photon carrying topological charge, $\Delta[\delta z] = \delta z_{m>0} - \delta z_{m=0}$. The data is fit using the full theoretical description of the group velocity for Bessel-Gauss modes Eq.~\eqref{eq:delay} (solid red line) and the ray tracing approach using Eq.~\eqref{eq:dDelta} (dashed blue line). We show a transverse profile of the $m=5$ beam intensity measured using a continuous wave source of the same wavelength at the second SLM, to show the effect of rotating the beam. A wedge is removed from the ring to illustrate that adding OAM rotates the intensity profile by an angle measured to be $\alpha = 155^{\circ}$ for all values of $m$. We note that although there is no dependence of the rotation angle on $m$, we find an increase in the delay for larger values of OAM due to the larger diameter of the ring aperture used, which is given by $D = 2\omega\sqrt{|m|/2}$.}
\label{results}}
\end{figure}
In order to better observe this delay we show the difference between the photon delays in Fig.~\ref{results}, i.e. the difference between the two sets of data in Fig.~\ref{fig:raw}, $\Delta[\delta z]=\delta z_{m>0}-\delta z_{m=0}$. Notwithstanding the uncertainty in the data that is related to the extremely small delays measured (at the very limit of what is currently achievable with HOM interferometry), all points consistently indicate a speed-up of the photons carrying OAM compared to those photons without. The solid red line is the same theoretical curve, Eq.~\eqref{eq:delay}, as described above. The dashed line shows the expected delays calculated from the simpler ray-tracing theory. This ray-tracing approach requires precise knowledge of the Poynting vector rotation angle, $\alpha$. We thus performed additional measurements with a standard diode laser injected into the arm containing the SLMs. We then impose the same phase and amplitude masks as for the single photon experiments but now with a 45 degree slice removed around $\alpha=0^\circ$.
The inset to Fig.~\ref{results} show just one example: for $m = 0$, the `slice' is rotated by $\alpha=180^\circ$, corresponding to the expected image inversion for our telescopic imaging setup, whereas for $m=\pm5$ the slice is only rotated by $\alpha=\pm155^\circ$. We verified that this rotation is independent of the OAM winding number. We therefore use this value of $\alpha$ in Eq.~\eqref{eq:dDelta}. The resulting theoretical prediction from this ray-tracing approach (with no free parameters) is shown as the dashed blue line in Fig.~\ref{results} and is in  good agreement with our experimental data.\\
\noindent {\bf Conclusions.} We have shown that the introduction of OAM onto a photon with a given spatial profile that is focused through a telescope, reduces the transit time over a fixed path length. This result is not in contradiction with previous results where a classical light beam spatial profile was allowed to vary as OAM was introduced. The main point here is that we are investigating the intrinsic OAM delay, rather than the overall effect due to OAM {\emph{and spatial reshaping} of the beam due to the OAM. This acceleration of the photon due to OAM can be understood within the framework of ray optics and is related to the fact that the longest path for a ray is given by rays that invert an image through a telescope. Beams or photons with OAM rotate images by an angle that is always less than $180^\circ$ and hence, energy flows along shorter paths. We also show that a more rigorous description can be obtained by explicitly calculating the group velocity (see Eq.~\eqref{eq:vg}), which when evaluated for a Bessel-Gauss beam carrying OAM provides a good match to the experimental data. Finally, we note that the intrinsic OAM photon-acceleration measured here can be viewed as an example of orbit-orbit coupling for photons \cite{spin-review}, in the sense that the {\emph orbital} angular momentum couples to the photon (path) \emph{orbit}.\\
\acknowledgements D. F. acknowledges financial support from the European Research Council under the European Union Seventh Framework Programme (FP/2007-2013)/ERC GA 306559 and EPSRC (UK, Grant No. EP/M01326X/1, EP/M006514/1). MJP D. F. acknowledges financial support from the European Research Council under the European Union Seventh Framework Programme ERC GA 340507. NW and CM acknowledge support from EPSRC CM-CDT Grant No. EP/L015110/1.

\noindent {\bf{Data availability.}}  All experimental data is available online

\end{document}